\documentclass[11pt]{article}

\usepackage[utf8]{inputenc}
\usepackage[T1]{fontenc}
\usepackage{amsmath,amssymb,amsfonts}
\usepackage{graphicx}
\usepackage[margin=1in]{geometry}
\usepackage{authblk}
\usepackage[colorlinks=true,citecolor=blue,linkcolor=blue,urlcolor=blue]{hyperref}
\usepackage[round]{natbib}
\usepackage{setspace}
\usepackage{booktabs}
\usepackage{caption}

\title{\bf Pattern-Based Sequential Multiple Imputation for Missing Data in Clinical Trials: An Extension for Baseline-Only Early Dropout Subjects}

\author[1]{Chen Zhang}
\author[2]{Junyu Nie}
\author[2]{Kexuan Li}
\author[2]{Ning Ding}

\affil[1]{Department of Epidemiology and Biostatistics, Arnold School of Public Health, University of South Carolina, Columbia, SC 29208, USA}
\affil[2]{Global Data and Quantitative Sciences, Bristol Myers Squibb, Princeton, NJ, USA}

\date{}

\begin{document}

\maketitle

\begin{center}
\textit{Corresponding author: Chen Zhang, Department of Epidemiology and Biostatistics, Arnold School of Public Health, University of South Carolina, Columbia, SC 29208, USA.} \\
\texttt{cz5@email.sc.edu}
\end{center}

\begin{abstract}
\noindent \textbf{Background:} The ICH E9 (R1) addendum establishes that treatment policy strategies for handling intercurrent events should target the treatment effect on a longitudinal continuous endpoint regardless of treatment discontinuation. Sequential multiple imputation (MI) models that condition each visit's imputation on discontinuation status or pattern reduce bias relative to mixed models for repeated measures and standard MI under a common missing-at-random assumption, but presuppose every subject contributes at least one post-baseline observation---an assumption violated by subjects who withdraw before any post-baseline assessment, a ``baseline-only early dropout'' pattern common in chronic-disease trials.

\textbf{Methods:} We propose Extended Pattern-based Sequential Multiple Imputation (EPSMI), which reconstructs missing data for baseline-only early dropouts using covariate-matched, same-arm donors before applying an extended discontinuation-pattern indicator within all eight existing sequential MI models. Two reconstruction strategies were evaluated: EPSMI-Full, which imputes the entire post-baseline trajectory from a donor, and EPSMI-Y1, which imputes only the first post-baseline visit, leaving subsequent visits to the pattern-extended sequential MI engine. A simulation study, grounded in the design and outcome characteristics of published trials in primary Sj\"{o}gren's syndrome, evaluated bias, coverage, precision, power, and Type I error across 24 scenarios combining early-dropout, off-treatment, discontinuation, and withdrawal mechanisms, comparing EPSMI against MMRM, standard MI, and the original sequential MI models applied after excluding early dropouts (No Early).

\textbf{Results:} Under random early dropout, both EPSMI strategies reduced bias relative to No Early. Under informative early dropout, the two strategies diverged: EPSMI-Y1 remained consistently robust, matching or exceeding No Early coverage (e.g., 87.2\% vs.\ 87.0\% overall for pattern-based models) with only mild Type I error inflation close to the nominal 5\% level, whereas EPSMI-Full's deterministic, single-step reconstruction produced larger bias, narrower confidence intervals driven by underestimated between-imputation variance, coverage below No Early in every model examined, and clear Type I error inflation (up to approximately 1.5 times nominal).

\textbf{Conclusion:} EPSMI-Y1 combined with pattern-based extended models is recommended as the primary analysis strategy for trials with baseline-only early dropout, allowing estimation to remain faithful to the treatment policy estimand over the full randomized population without an implicitly narrowed target population; EPSMI-Full's less reliable performance restricts its role to a simple sensitivity analysis.

\vspace{0.5em}
\noindent \textbf{Keywords:} treatment policy estimand, missing data, multiple imputation, early dropout, donor imputation, sequential imputation
\end{abstract}

\vspace{1em}
\noindent \textbf{Abbreviations:} ANCOVA, analysis of covariance; CI, confidence interval; DAR, discontinuation at random; DNAR, discontinuation not at random; EDAR, early dropout at random; EDNAR, early dropout not at random; EPSMI, Extended Pattern-based Sequential Multiple Imputation; ESSDAI, EULAR Sj\"{o}gren's Syndrome Disease Activity Index; ICH, International Council for Harmonisation; MAR, missing at random; MI, multiple imputation; MICE, multiple imputation by chained equations; MMRM, mixed model for repeated measures; pSS, primary Sj\"{o}gren's syndrome; RTB, return-to-baseline; SAA, stay-as-assigned.

\section{Introduction}\label{sec1}

The ICH E9 (R1) addendum on estimands and sensitivity analysis formalized the requirement that clinical trials pre-specify how post-baseline intercurrent events, such as treatment discontinuation, are to be handled in defining the treatment effect of interest \citep{ich_e9r1_2019}. For continuous, repeatedly measured endpoints, the treatment policy strategy targets the effect of treatment regardless of whether patients remain on assigned therapy. This approach is now common in confirmatory trials for chronic conditions, where outcomes collected after discontinuation, i.e., off-treatment data, are as relevant to the estimand as on-treatment outcomes. Concretely, the treatment effect targeted under this strategy is the between-arm difference in mean change from baseline to the final assessment visit, computed over the full randomized population.

Estimation of such estimands has traditionally relied on either MMRM fitted to all observed pre- and post-discontinuation data, or multiple imputation (MI) pooled via Rubin's rules \citep{rubin_multiple_2004,bell_mixed_2020}. Both rely on a common missing-at-random (MAR) assumption: missing outcomes resemble those of patients with similar observed history, irrespective of discontinuation status. This assumption becomes implausible when on- and off-treatment trajectories differ and missingness is more likely after discontinuation \citep{jin_handling_2026}. To address this, a family of sequential MI models has been proposed that conditions each visit's imputation on a patient's discontinuation status or pattern, in addition to prior outcomes \citep{drury_estimation_2024}. These models range from the simplest (CICS, which ignores treatment status) to increasingly flexible variants with separate intercepts and/or slopes by on/off-treatment status (OICS, OIOS) or full discontinuation pattern (PICS, PIOS, PIPS), plus residual-conditioned analogues (OICS-R, PICS-R), and were shown to reduce bias relative to MMRM and pooled MI at the cost of increased variance with greater flexibility. This work provides the methodological foundation for the present paper.

A structural limitation, however, is that the sequential MI framework presupposes every subject contributes at least one post-baseline outcome, so that a discontinuation indicator or pattern can be defined. In practice, some patients are lost to the study before any post-baseline assessment is collected. Reasons include withdrawal of consent, loss to follow-up, or other causes unrelated to a documented discontinuation event. For patients who contributed only a baseline assessment before withdrawal, no discontinuation-time or pattern indicator can be meaningfully defined, and applying the existing models as specified requires either excluding them or absorbing them into an ill-fitting undetermined category that fails to capture the distinct nature of their missingness.

This pattern is not uncommon in practice, and regulatory guidance further motivates addressing it directly rather than through exclusion. Baseline-only early dropout has been observed across a range of confirmatory trial settings \citep{ibrahim_characterization_2022}. Both the ICH E9 (R1) addendum and related regulatory guidance on missing data favor primary analyses conducted on the full randomized population, and generally discourage modified intention-to-treat populations that exclude subjects lacking any post-baseline assessment, since such exclusions can bias the estimated treatment effect toward a non-randomized subpopulation \citep{ich_e9r1_2019,mallinckrodt_handling_2026,mallinckrodt_aligning_2020,national_research_council_us_panel_on_handling_missing_data_in_clinical_trials_prevention_2010}.

This issue is not merely theoretical. Randomized trials in primary Sj\"{o}gren's syndrome (pSS), a chronic autoimmune disease assessed longitudinally via the ESSDAI index, illustrate its practical relevance \citep{seror_eular_2010}. Published phase II-b and phase III trials of B-cell- and T-cell-targeted biologics in moderate-to-severe pSS report meaningful proportions of patients withdrawing early, for reasons including adverse events, non-compliance, and patient decision \citep{bowman_safety_2022,baer_efficacy_2021}. Alongside this early withdrawal, these trials show substantial heterogeneity in ESSDAI trajectories, with one trial reporting a pronounced placebo response that obscured an otherwise biologically active treatment. These trials motivate the clinical relevance of the problem and the specific parameterization used to ground our simulation in a realistic disease and trial context.

We propose the Extended Pattern-based Sequential Multiple Imputation (EPSMI) framework, which resolves this gap by borrowing post-baseline information from covariate-matched, non-early-dropout donors within the same randomized arm---an approach related to hot-deck donor imputation in the survey literature \citep{fuller_hot_2005,andridge_review_2010}, and then extending the sequential MI machinery to accommodate the reconstructed subjects. Specifically, EPSMI (i) constructs, for each baseline-only early dropout subject, a pool of eligible same-arm donors matched on standardized baseline covariates, (ii) reconstructs either the subject's entire post-baseline trajectory (EPSMI-Full) or only the first post-baseline visit (EPSMI-Y1) from a selected donor, and (iii) applies an extended version of each of the eight sequential MI models, in which the discontinuation pattern indicator is augmented with an explicit early dropout category, to impute any remaining missing values and distinguish reconstructed early-dropout subjects from conventional treatment discontinuers. The remainder of this paper is organized as follows. Section 2 details the notation, existing sequential MI models, and the EPSMI extension. Section 3 describes a simulation study, grounded in the design and outcome characteristics of the ianalumab and abatacept pSS trials, comparing EPSMI against MMRM, standard MI, and the original sequential MI models restricted to non-early-dropout subjects. Section 4 presents results for bias, coverage, precision, and Type I error. Section 5 concludes with recommendations and discussion of limitations.

\section{Methods}\label{sec2}

\subsection{Notation and Estimand}\label{sec2-1}

Consider a two-arm, parallel-group randomized trial with $Z_i \in \{0,1\}$ denoting the randomized treatment assignment (control, active) for subject $i$. Outcomes $Y_{ij}$ are measured at visits $j = 0,\dots,J$, with $j=0$ denoting baseline and $J$ the final assessment. As in the underlying sequential MI framework \citep{drury_estimation_2024}, let $D_{ij}$ indicate whether subject $i$ remains on randomized treatment ($D_{ij}=0$) or has discontinued ($D_{ij}=1$) by visit $j$, assumed monotone, and let $P_{ij}$ denote the categorical discontinuation pattern up to visit $j$.

We additionally define $E_i \in \{0,1\}$, an indicator of baseline-only early dropout: $E_i = 1$ if subject $i$ has $Y_{i0}$ observed but all post-baseline values missing, with no documented treatment discontinuation event. This is treated as structurally distinct from $D_{ij}$, since it reflects withdrawal before any post-baseline assessment rather than an observed on-to-off-treatment switch; $D_{ij}$ and $P_{ij}$ are consequently undefined for $j\ge1$ whenever $E_i=1$.

The estimand of interest, consistent with the treatment policy strategy, is
\begin{equation}
\Delta_J = E\left[Y_{iJ} - Y_{i0} \mid Z_i = 1\right] - E\left[Y_{iJ} - Y_{i0} \mid Z_i = 0\right],
\label{eq:estimand}
\end{equation}
the difference in expected change from baseline to the final visit between the active and control arms, regardless of treatment discontinuation, early dropout, or subsequent study withdrawal.

\subsection{Common-MAR Estimation Approaches}\label{sec2-2}

Two standard estimation strategies, relying on a common MAR assumption that does not distinguish the missingness mechanism, serve as benchmarks. MMRM analyzes all observed pre- and post-discontinuation data directly via a repeated-measures mixed model with treatment-by-visit and baseline-by-visit interactions and a subject-level random intercept, with the treatment effect at the final visit extracted by re-parameterizing the visit factor \citep{bell_mixed_2020}:
\begin{equation}
Y = \text{Intercept} + Z + \text{Timepoint} + Y_0 + Z \times \text{Timepoint} + Y_0 \times \text{Timepoint}.
\label{eq:mmrm}
\end{equation}
Standard MI (MICE) imputes missing post-baseline values jointly by chained equations with predictive mean matching, using treatment arm and baseline covariates as predictors, without any term distinguishing on- from off-treatment status \citep{azur_multiple_2011}:
\begin{equation}
Y_j = \text{Intercept} + Z + Y_0, \quad j = 1,\dots,J.
\label{eq:mice}
\end{equation}

\subsection{Sequential MI models Using Previous Outcomes}\label{sec2-3}

The sequential MI framework \citep{drury_estimation_2024} imputes each missing visit $j = 1,\dots,J$ in turn, separately by treatment arm, using a regression on previously observed or imputed outcomes $Y_0,\dots,Y_{j-1}$, augmented with terms reflecting on/off-treatment status or discontinuation pattern:

\noindent\textbf{Common Intercepts Common Slopes (CICS):}
\begin{equation}
Y_j = \text{Intercept} + Y_0 + \cdots + Y_{j-1}.
\end{equation}

\noindent\textbf{On/Off-Intercepts Common Slopes (OICS):}
\begin{equation}
Y_j = \text{Intercept} + D_j + Y_0 + \cdots + Y_{j-1}.
\end{equation}

\noindent\textbf{Pattern Intercepts Common Slopes (PICS):}
\begin{equation}
Y_j = \text{Intercept} + P_j + Y_0 + \cdots + Y_{j-1}.
\end{equation}

\noindent\textbf{On/Off-Intercepts On/Off-Slopes (OIOS):}
\begin{equation}
Y_j = \text{Intercept} + D_j + Y_0 + \cdots + Y_{j-1} + D_j * Y_0 + \cdots + D_j * Y_{j-1}.
\end{equation}

\noindent\textbf{Pattern Intercepts On/Off-Slopes (PIOS):}
\begin{equation}
Y_j = \text{Intercept} + P_j + Y_0 + \cdots + Y_{j-1} + D_0 * Y_0 + \cdots + D_{j-1} * Y_{j-1}.
\end{equation}

\noindent\textbf{Pattern Intercepts Pattern Slopes (PIPS):}
\begin{equation}
Y_j = \text{Intercept} + P_j + Y_0 + \cdots + Y_{j-1} + P_j * Y_1 + \cdots + P_j * Y_{j-1}.
\end{equation}

Two residual-based analogues condition on residuals $R_j = Y_j - \hat{\mu}_j$ from the previous step rather than raw outcomes:

\noindent\textbf{On/Off-Intercepts With Common Slopes Using Residuals (OICS-R):}
\begin{equation}
Y_j = \text{Intercept} + D_j + R_0 + R_1 + \cdots + R_{j-1}.
\end{equation}

\noindent\textbf{Pattern Intercepts With Common Slopes Using Residuals (PICS-R):}
\begin{equation}
Y_j = \text{Intercept} + P_j + R_0 + R_1 + \cdots + R_{j-1}.
\end{equation}

Each model is fitted separately within each of $m$ imputations and each randomized arm; a Gaussian draw around the fitted conditional mean is used to impute missing values, and completed datasets are analyzed with the ANCOVA model described in Section 2.5.

\subsection{Baseline-only Early Dropout and the EPSMI Extension}\label{sec2-4}

The models of Section 2.3 presuppose that every subject contributes at least one post-baseline outcome, so that $D_j$ or $P_j$ can be defined. For subjects with $E_i=1$, this fails, and three unsatisfactory options arise if the models are applied without modification: excluding these subjects entirely (which discards baseline information and, whenever early dropout is informative, biases the estimand toward a systematically selected subpopulation); assigning a default status to $D_j/P_j$ (which is factually incorrect and contaminates the sequential regressions for all patients); or imputing $Y_1$ from $Y_0$ alone via CICS-type conditioning (which ignores realized post-baseline trajectories of similar, arm-matched patients and provides no basis for distinguishing early dropouts from other patients at later visits).

To resolve this, we propose the Extended Pattern-based Sequential Multiple Imputation (EPSMI) framework, which combines a donor-based reconstruction step with an explicit extension of the discontinuation-pattern indicator to an ``early dropout'' category, allowing all eight sequential MI models of Section 2.3 to be applied to the full randomized sample.

\textbf{Donor pool construction.} For each subject $i$ with $E_i=1$, a donor pool is constructed from eligible subjects $i'$ in the same randomized arm with $E_{i'}=0$ and complete data on the outcome(s) required for the chosen reconstruction strategy. Standardized baseline covariates are used to compute the Euclidean distance between subject $i$ and each candidate donor; donors within a pre-specified caliper are retained, with the nearest available donor(s) used as a fallback if none fall within the caliper. A single donor is then selected either by nearest-neighbor matching or random selection from within the caliper-restricted pool.

\textbf{EPSMI-Full.} The donor's entire post-baseline trajectory is copied to the early-dropout subject, who is thereafter treated as a fully observed on-treatment case unless further, non-structural missingness requires imputation.

\textbf{EPSMI-Y1.} Only the first post-baseline visit is copied from the donor. The pattern variable $P_j$ is extended so that, at each visit, early-dropout subjects carry an additional ``Early'' level alongside the usual discontinuation-time levels, and this extended $P_j$ is used exactly as in Section 2.3 to impute the remaining visits, giving early-dropout subjects their own intercept (and, where applicable, slope) terms within the pattern-based models.

Applying each of the eight models of Section 2.3 to the reconstructed, pattern-extended dataset under either strategy yields eight corresponding extended models, denoted with an ``E'' prefix: ECICS, EOICS, EPICS, EOIOS, EPIOS, EPIPS, EOICS-R, and EPICS-R.

\subsection{Final Analysis Model and Pooling}\label{sec2-5}

For every method that yields a completed dataset (MICE, the sequential MI models of Section 2.3, and the EPSMI-extended models of Section 2.4), the same ANCOVA model is applied to each completed dataset:
\begin{equation}
Y_{iJ} - Y_{i0} = \beta_0 + \beta_1 Z_i + \beta_2 Y_{i0} + \varepsilon_i,
\label{eq:ancova}
\end{equation}
with $\widehat{\beta}_1$ the estimated treatment effect. MMRM instead estimates the treatment effect directly from the mixed model fitted to the observed data, without a separate completed-data analysis step (Section 2.2).

For the completed-dataset methods, the resulting treatment-effect estimates and variances from each of the $m$ imputations are pooled using Rubin's rules, yielding a pooled point estimate, standard error, and degrees of freedom used to construct confidence intervals and p-values \citep{rubin_multiple_2012,rubin_multiple_2004,burgess_combining_2013}. MMRM, which does not produce multiple completed datasets, requires no such pooling step.

\section{Simulation}\label{sec3}

The goal of the simulation study was to evaluate the bias, precision, coverage, and Type I error of EPSMI relative to MMRM, standard MI, and the original sequential MI models restricted to non-early-dropout subjects, across a range of plausible early-dropout, discontinuation, and withdrawal scenarios. The design of the simulated trials matched the two-arm parallel-group structure of the pSS trials motivating this work, with 300 patients per arm. For each simulated trial, we considered two early-dropout mechanisms, two treatment-discontinuation mechanisms, two off-treatment expected-outcome scenarios, and three study-withdrawal profiles, resulting in $2\times2\times2\times3=24$ scenarios, each replicated 1000 times under a treatment-effect data-generating mechanism and, separately, under a null mechanism for evaluating Type I error.

\subsection{Data Generation}\label{sec3-1}

To create outcome data for virtual patients, we generated a vector of correlated patient-level data comprising potential on-treatment and off-treatment ESSDAI values for all visits, using baseline covariates (age, disease duration, sex) and a common four-dimensional correlated latent process, with subject-level heterogeneity added separately to the on- and off-treatment trajectories. The expected values of ESSDAI for baseline and on-treatment outcomes in the control arm were set as $(9.4, 8.8, 8.1, 7.5)$, and the on-treatment treatment effect (difference between active and control) was set to $(0, -3, -3, -3)$ for visits $0$--$3$, respectively, reflecting baseline disease activity levels and a clinically meaningful post-baseline reduction consistent with published pSS trials (Section 1). All generated values were bounded to the valid ESSDAI range. Full parameterization details of the data generation process are given in the eFiles.

\subsection{Missingness Mechanisms}\label{sec3-2}

Within each arm, a fixed proportion $p^{ED}=10\%$ of subjects was designated as baseline-only early dropouts ($E_i=1$; all post-baseline values missing). Two mechanisms were considered: \textbf{early dropout at random (EDAR)}, based on a uniform random score assigned independently of outcome, with the highest-scoring subjects in each arm selected as early dropouts; and \textbf{early dropout not at random (EDNAR)}, based on a prognostic score linked to each subject's unobserved future on-treatment outcome, disease duration, and age, such that subjects with worse anticipated outcomes, longer disease duration, and younger age were more likely to be designated as early dropouts.

Among subjects not designated as early dropouts, two mechanisms were considered for treatment discontinuation, both based on lack of efficacy: \textbf{discontinuation at random (DAR)}, using a propensity score linked to each subject's previous on-treatment ESSDAI value, with the worst-ranked subjects discontinuing; and \textbf{discontinuation not at random (DNAR)}, using a propensity score linked to each subject's next potential on-treatment ESSDAI value. A fixed discontinuation rate of 20\% per arm was allocated across the post-baseline visits in a 5:3:2 ratio, producing a larger proportion of discontinuations at earlier visits.

Two off-treatment expected-outcome scenarios were considered: \textbf{return-to-baseline (RTB)}, in which discontinuers' expected outcomes revert toward their pre-treatment (baseline) level following discontinuation, and \textbf{stay-as-assigned (SAA)}, in which discontinuers' expected outcomes continue to follow the trajectory associated with their randomized treatment. Full parameterization details for both scenarios are given in the eFiles.

Following discontinuation, study withdrawal (resulting in subsequent outcomes being set to missing) was simulated as a single assessment at the point of discontinuation, with the probability of withdrawal varied across three profiles: \textbf{more early withdrawal (EW)}, larger withdrawal probability for subjects discontinuing at the first post-baseline visit; \textbf{balanced withdrawal (BW)}, equal withdrawal probability regardless of discontinuation timing; and \textbf{more late withdrawal (LW)}, larger withdrawal probability for subjects discontinuing at later visits.

\subsection{Analysis and Performance Measures}\label{sec3-3}

Within each simulated trial, the treatment effect was estimated using MMRM, standard MI, the eight original sequential MI models applied after excluding early-dropout subjects (No Early), and the eight extended sequential MI models under both EPSMI-Full and EPSMI-Y1 reconstruction, each with random and nearest-neighbor donor selection. All MI-based procedures used $m=25$ imputations, and the analysis model applied to each completed dataset was the ANCOVA model of Section 2.5, with estimates pooled via Rubin's rules.

The performance measures used for each method were the bias of the estimated treatment effect relative to the analytically consistent treatment effect derived from the complete (pre-missingness) simulated data, the 95\% confidence interval coverage probability, the precision of the estimated treatment effect (mean 95\% CI halfwidth), power, and type I error, evaluated under the null data-generating mechanism.

\section{Results}\label{sec4}

\subsection{Bias}\label{sec4-1}

Bias is compared across four method categories: the common-MAR benchmarks (MMRM, MICE), the original sequential MI models applied after excluding early-dropout subjects (No Early), and the two EPSMI reconstruction strategies (Full Donor, Y1 Donor) applied to the extended sequential MI models. We first check whether bias responds to the data-generating mechanism dimensions in the expected directions, as a consistency check on the simulation design, before turning to the central comparison of interest: the relative performance of No Early, EPSMI-Full, and EPSMI-Y1.

\subsubsection{Consistency Checks on the Data-Generating Mechanisms}\label{sec4-1-1}

\textbf{EDAR vs.\ EDNAR.} Bias is consistently larger under EDNAR than under EDAR across all method categories, including the benchmarks, No Early, and both EPSMI strategies (Figure 1). This is expected: under EDNAR, early dropout depends on a subject's unobserved future prognosis, disease duration, and age, so early-dropout subjects are systematically distinct in outcome distribution from the remaining sample---whether that sample is used to exclude them (No Early) or to construct a donor pool (EPSMI). Under EDAR, early dropout is independent of outcome, so the mechanism introduces no systematic distortion regardless of how it is handled.

\begin{figure}[t]
\centering
\includegraphics[width=\textwidth]{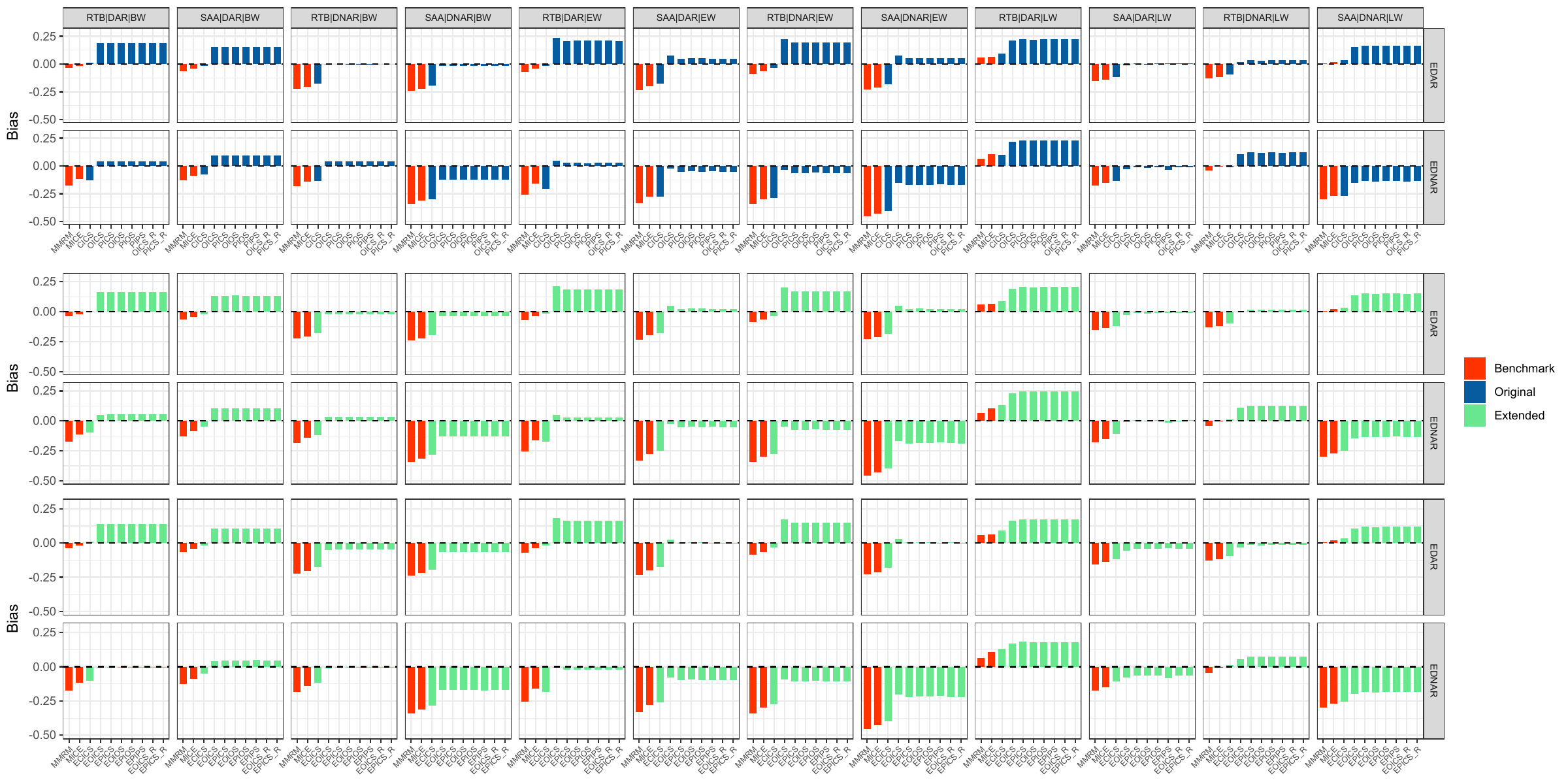}
\caption{Bias of the estimated treatment effect across all 24 simulation scenarios.}
\label{Figure1}
\end{figure}

\textbf{RTB vs.\ SAA.} Benchmark and pattern-based MI methods show opposite sensitivity to the off-treatment trajectory assumption (Figure 1). For MMRM and MICE, bias is consistently larger under SAA than under RTB: under SAA, discontinuers remain on the active-treatment trajectory, making observed data more selectively favorable and requiring greater extrapolation under the common-MAR assumption; under RTB, discontinuers' outcomes revert toward placebo-like values, narrowing the observed--missing gap. For pattern-based sequential MI (PICS, PICS-R, and their extended counterparts), the pattern reverses: bias is larger under RTB than under SAA. Under RTB, the true treatment effect itself dilutes over time as discontinuers converge toward the placebo trajectory, so the models must track a non-constant target; under SAA, the treatment effect remains stable across visits and is more easily learned and extrapolated.

\textbf{DAR vs.\ DNAR.} Bias is generally larger under DNAR than under DAR, most clearly for the benchmark methods (Figure 1). Under DNAR, discontinuation depends on a subject's future, unobserved potential outcome, so subjects with poorer latent prognosis discontinue before that outcome is ever observed, producing informative missingness that no method can fully explain from observed data alone. Under DAR, discontinuation depends only on already-observed history, which the models can condition on directly. This gap is not uniform across scenarios. Three factors attenuate it: RTB introduces treatment dilution that compresses the scope for DAR/DNAR to diverge; higher withdrawal rates reduce the post-discontinuation information available to distinguish the two mechanisms; and the sequential MI models' conditioning on baseline, prior outcomes, and pattern variables absorbs part of the discontinuation-mechanism signal regardless of type. Overall, DNAR remains more challenging than DAR, particularly for benchmark methods, but the magnitude depends jointly on the off-treatment assumption and withdrawal profile.

\subsubsection{No Early vs EPSMI-Full vs EPSMI-Y1}\label{sec4-1-2}

Under EDAR, both EPSMI strategies improve upon No Early. The bias difference $|\mathrm{Bias}(\text{Extended})| - |\mathrm{Bias}(\text{Original})|$ is negative across nearly all scenarios for both Full Donor and Y1 Donor, under both donor-selection rules (Figure 2), with the largest gains for PICS, PIPS, and PICS-R. This finding aligns with the design of EDAR, since the donor pool under EDAR is not systematically different from early-dropout subjects, so reconstructing either the full trajectory or only the first visit recovers largely unbiased information.

Under EDNAR, the two strategies diverge sharply. Y1 Donor continues to improve on No Early in nearly every scenario, with the bias difference remaining negative or near zero across the range of off-treatment, discontinuation, and withdrawal combinations examined (Figure 2). Full Donor becomes unreliable: in many scenarios the bias difference turns positive, indicating it introduces more bias than simple exclusion. This arises because early-dropout subjects under EDNAR differ systematically in prognosis from the donor pool by construction, and Full Donor's deterministic, single-step copy of an entire trajectory carries any donor--recipient mismatch forward across all three post-baseline visits. Y1 Donor instead borrows only the donor's first, most reliably matched value, while the extended pattern-based models impute later visits using the explicit early pattern level together with observed history-outcome relationships across the broader sample---absorbing the systematic donor mismatch statistically rather than through a single substitution.

\begin{figure}[t]
\centering
\includegraphics[width=\textwidth]{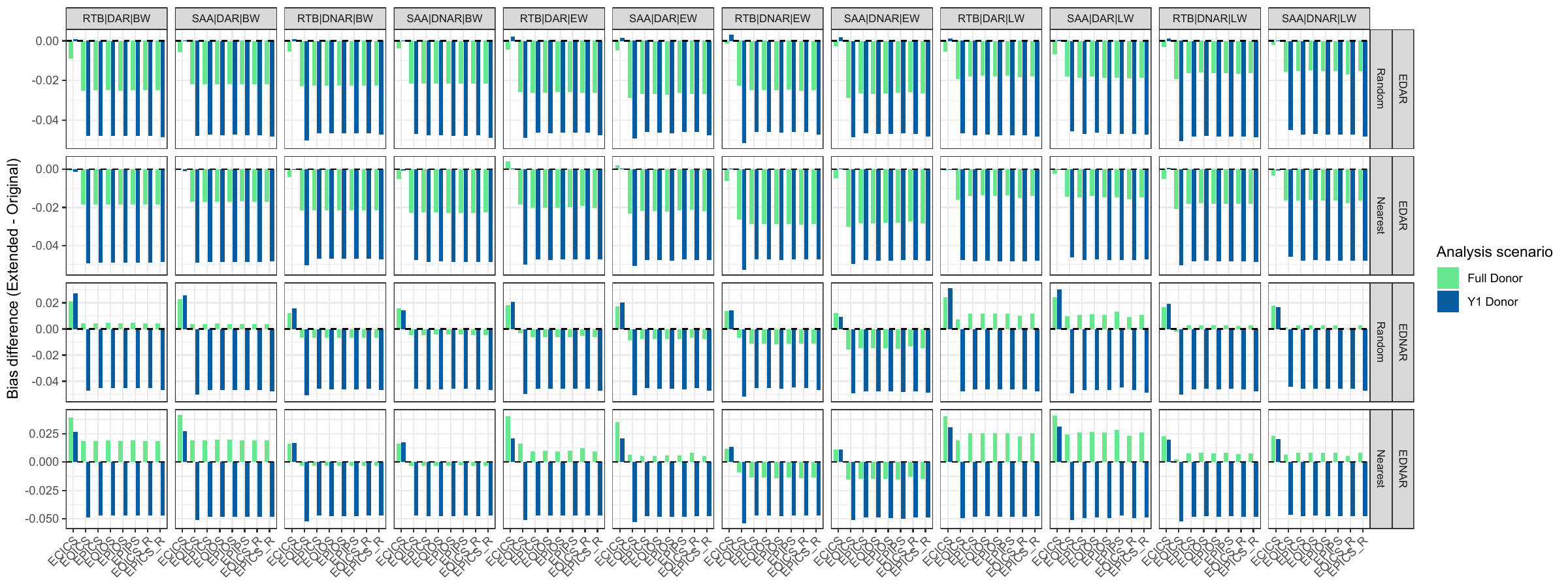}
\caption{Bias difference (Extended $-$ Original) for each of the eight EPSMI models across all 24 simulation scenarios.}
\label{Figure2}
\end{figure}

Taken together, \textbf{Y1 Donor is the most consistently strong-performing EPSMI strategy}: it matches or improves upon No Early under both EDAR and EDNAR, and outperforms Full Donor specifically where accurate handling of early dropout matters most. Full Donor is reliable only under EDAR and is not a robust general-purpose strategy.

Donor selection affects the two EPSMI strategies differently (Figure 1, Figure 2). For Full Donor, nearest-neighbor selection produces larger and more frequent positive bias differences under EDNAR than random selection, since deterministically matching the single nearest donor on observed covariates tends to lock in a covariate-plausible but outcome-mismatched substitution, whereas random selection allows this mismatch to be partially averaged out. For Y1 Donor, bias performance is essentially unaffected by the choice of donor-selection rule, remaining favorable relative to No Early under both. Across the three withdrawal profiles examined, the bias advantage of Y1 Donor and its pattern-based extensions holds consistently, indicating it is not specific to a particular withdrawal timing pattern.

\subsection{Other Performance Metrics}\label{sec4-2}

Table 1 (coverage), Table 2 (precision/95\% halfwidth), Table 3 (power), and Table 4 (Type I error) summarize performance averaged across the 24 scenarios (or, for Type I error, the corresponding 24 scenarios under the null data-generating mechanism), with results further averaged within each scenario dimension (EDAR/EDNAR, RTB/SAA, DAR/DNAR, EW/BW/LW) to form the Overall column and dimension-specific columns. For clarity, Tables 1--4 focus on CICS and the three best-performing pattern-based models identified in Section 4.1 (PICS, PIPS, PICS-R), alongside the MMRM and MICE benchmarks. Unlike Figure 1, which pools the two donor-reconstruction strategies, Tables 1--4 report the Extended models separately by Full Donor and Y1 Donor reconstruction, allowing the two strategies to be compared directly against each other and against the corresponding No Early models.

\subsubsection{95\% CI Coverage}\label{sec4-2-1}

MMRM shows the lowest coverage overall (0.577), particularly under SAA (0.438) and EDNAR (0.441) (Table 1). MICE (0.717) and No Early CICS (0.723) show intermediate coverage, while the No Early pattern-based models (PICS, PIPS, PICS-R, each 0.870--0.871 overall) achieve markedly higher coverage but remain below the nominal 95\% level.

Separating the two Extended strategies shows they behave quite differently relative to No Early. Y1 Donor matches or slightly exceeds the corresponding No Early pattern-based models overall (e.g., EPICS 0.872 vs.\ PICS 0.870; EPICS-R 0.873 vs.\ PICS-R 0.870), whereas Full Donor is uniformly lower than No Early across all four models examined (e.g., EPICS 0.847 vs.\ PICS 0.870). This overall advantage for Y1 Donor is driven mainly by EDAR, where Y1 Donor coverage for the pattern-based models rises to 0.906--0.908, several points above the corresponding No Early values (0.870--0.871), while Full Donor remains essentially flat (0.865). Under EDNAR, both Extended strategies show reduced coverage relative to No Early for the pattern-based models (No Early 0.871--0.873 vs.\ Full Donor 0.830--0.831 vs.\ Y1 Donor 0.838), consistent with the larger bias observed for both strategies in Section 4.1.2 when early dropout is informative; the shortfall is nonetheless smaller for Y1 Donor than for Full Donor in every pattern-based model. Reported jointly, as in Section 4.1's pooled figures, this would understate how much better Y1 Donor performs than Full Donor.

\begin{table}[t]
\centering
\caption{95\% CI coverage of the estimated treatment effect, averaged over all simulation scenarios and by scenario dimension (EDAR/EDNAR, RTB/SAA, DAR/DNAR, EW/BW/LW).}
\label{tab:Table1}
\includegraphics[width=0.95\textwidth]{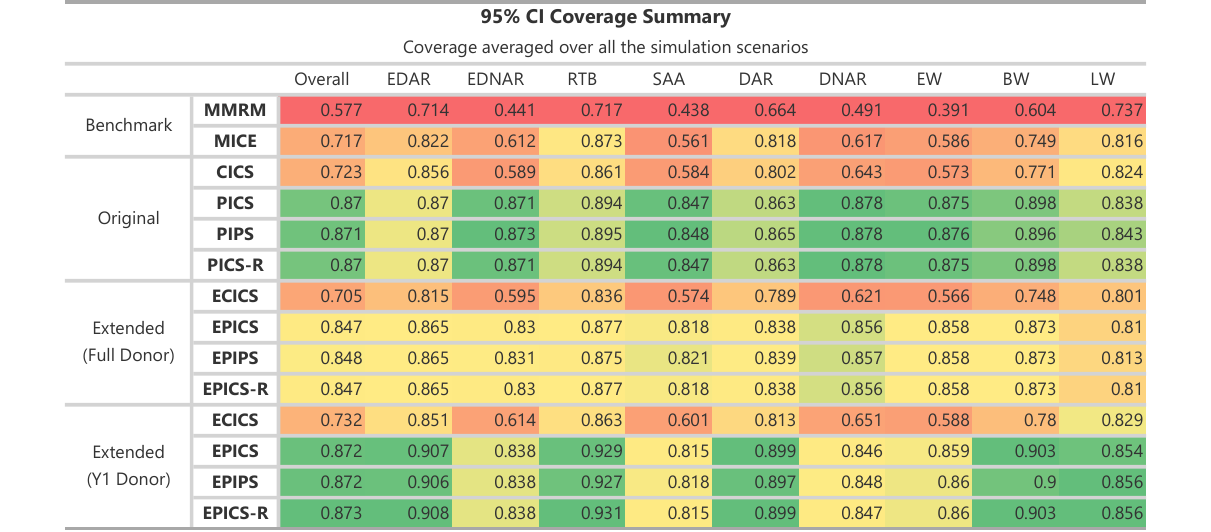}
\end{table}

\subsubsection{Precision (95\% CI Halfwidth)}\label{sec4-2-2}

Halfwidths are narrowest for MMRM (0.213 overall), followed by No Early CICS (0.246), with the No Early pattern-based models (0.266--0.275) and MICE (0.255) producing the widest intervals (Table 2). Both Extended strategies produce narrower halfwidths than their No Early counterparts, but Full Donor narrows the intervals more aggressively than Y1 Donor (e.g., for PICS: No Early 0.266 vs.\ Full Donor 0.252 vs.\ Y1 Donor 0.256).

This ordering is consistent with Full Donor's fully deterministic, one-time trajectory copy, which contributes essentially no between-imputation variance for reconstructed subjects, whereas Y1 Donor deterministically fixes only the first post-baseline visit and lets the pattern-extended sequential MI engine impute the remaining visits stochastically, retaining more of the appropriate uncertainty. Combined with Section 4.2.1, this indicates that under EDNAR, Full Donor's narrower intervals compound with its larger bias to produce the more severe coverage shortfall of the two strategies, while Y1 Donor's more modest narrowing better preserves calibration.

\begin{table}[t]
\centering
\caption{Precision (95\% CI halfwidth) of the estimated treatment effect, averaged over all simulation scenarios and by scenario dimension (EDAR/EDNAR, RTB/SAA, DAR/DNAR, EW/BW/LW).}
\label{tab:Table2}
\includegraphics[width=0.95\textwidth]{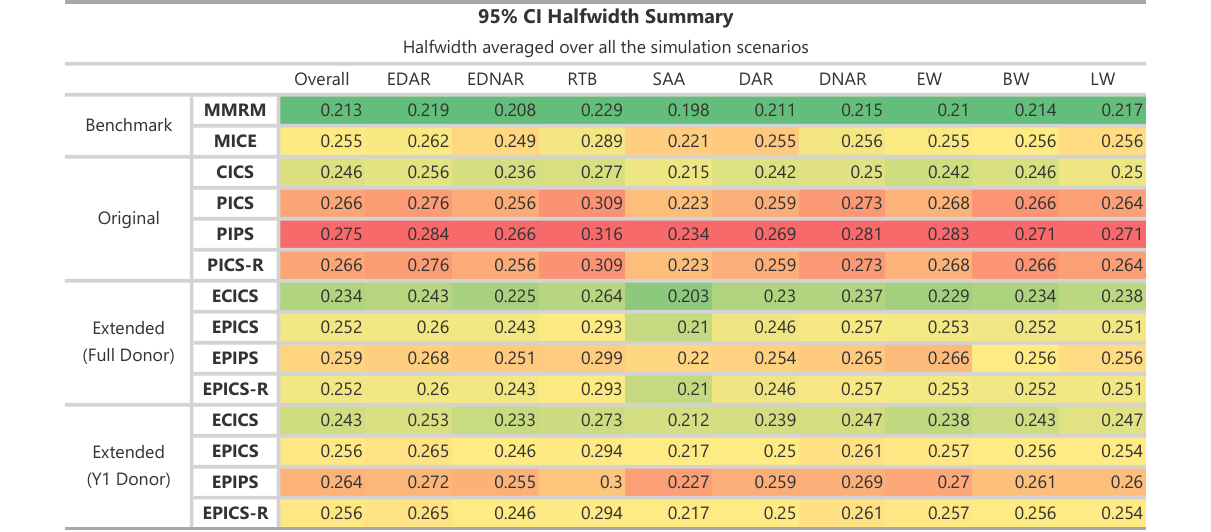}
\end{table}

\subsubsection{Power}\label{sec4-2-3}

Table 3 summarizes the power ($p<0.05$) to detect the simulated treatment effect. Power was uniformly high across all methods (0.93--0.99 overall), reflecting the large sample size and effect size used in the data-generating mechanism. MMRM (0.989) showed the highest overall power, followed by MICE (0.980) and No Early CICS (0.979); among the No Early pattern-based models, PIPS (0.934) was consistently lower than PICS and PICS-R (0.951), mirroring its greater estimation variance (Section 4.2.2).

Both Extended strategies showed a modest power gain over their No Early pattern-based counterparts (e.g., PICS: No Early 0.951 vs.\ Full Donor 0.961 vs.\ Y1 Donor 0.965), with Y1 Donor achieving marginally higher power than Full Donor for every pattern-based model. For CICS the pattern was flat to slightly reversed (No Early 0.979 vs.\ Full Donor 0.980 vs.\ Y1 Donor 0.977), indicating the gain is specific to the pattern-based models rather than a general feature of donor reconstruction.

MMRM's power advantage should not be read as a genuine detection advantage: as shown in Section 4.2.4, it coincides with substantially inflated Type I error, so power is not directly comparable across methods with differing degrees of miscalibration. The modest power gain for Y1 Donor is more defensible in this respect, since, unlike Full Donor, it is not accompanied by material Type I error inflation (Section 4.2.4), and is consistent with the larger effective sample size retained once early dropouts are reconstructed rather than excluded.

\begin{table}[t]
\centering
\caption{Power of the estimated treatment effect, averaged over all simulation scenarios and by scenario dimension (EDAR/EDNAR, RTB/SAA, DAR/DNAR, EW/BW/LW).}
\label{tab:Table3}
\includegraphics[width=0.95\textwidth]{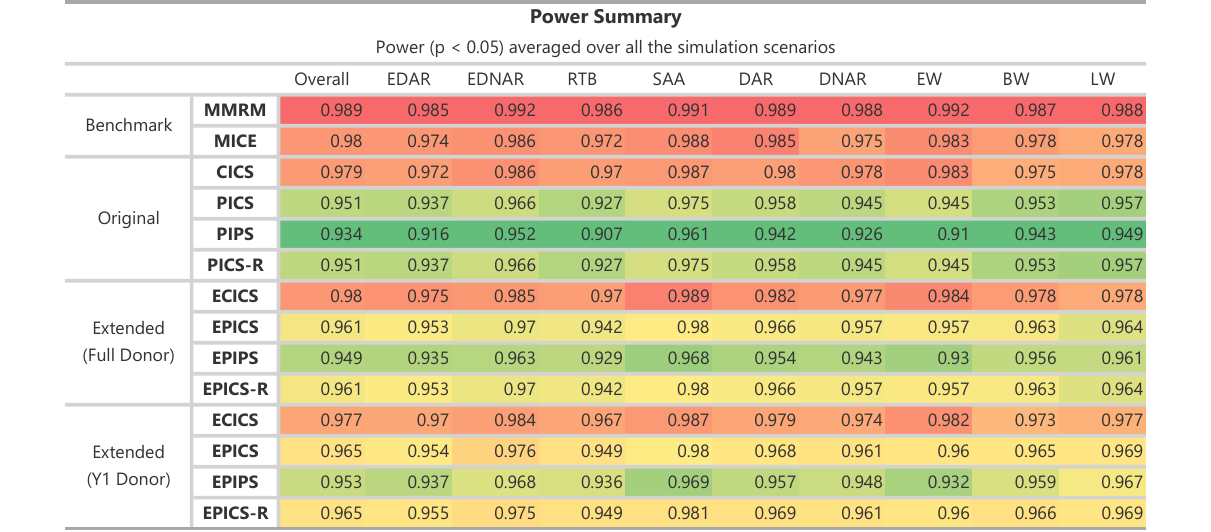}
\end{table}

\subsubsection{Type I Error}\label{sec4-2-4}

Under the null data-generating mechanism, MMRM shows inflated Type I error above the nominal 5\% level in every scenario dimension (0.071--0.080), while MICE remains closer to nominal (0.051--0.064) (Table 4). No Early pattern-based methods are generally well controlled, with PICS and PICS-R at 0.051 overall and PIPS at 0.059.

The two Extended strategies diverge sharply on this measure. Full Donor produces clear Type I error inflation across every model, roughly 1.4--1.5 times the corresponding No Early value overall (e.g., EPICS 0.075 vs.\ PICS 0.051; EPICS-R 0.075 vs.\ PICS-R 0.051; ECICS 0.082 vs.\ CICS 0.055), and rises further under SAA and EDNAR, reaching 0.095 (EPICS, EPICS-R) to 0.101 (EPIPS) under SAA. Y1 Donor, by contrast, remains close to nominal and close to the corresponding No Early values across every model and dimension (e.g., EPICS 0.055 vs.\ PICS 0.051 overall; EPICS 0.071 vs.\ PICS 0.068 under SAA), with only mild inflation, most visible under SAA and EDNAR, consistent with the bias and precision patterns of Sections 4.1.2 and 4.2.2.

Read together with Table 2, Full Donor's Type I error inflation is attributable predominantly to its narrower, variance-underestimated confidence intervals rather than to point-estimate bias, since its largest inflation (under SAA) does not coincide with its largest bias, which instead peaks under RTB (Section 4.1.2); Y1 Donor's much smaller inflation is consistent with its more modest interval narrowing (Section 4.2.2). Because the tables are now broken down by reconstruction strategy, this confirms directly, rather than merely inferring, that Full Donor's variance underestimation---not Y1 Donor's---is the primary driver of the Type I error concerns raised for the Extended models as a whole.

\begin{table}[t]
\centering
\caption{Type I error of the estimated treatment effect, averaged over all simulation scenarios and by scenario dimension (EDAR/EDNAR, RTB/SAA, DAR/DNAR, EW/BW/LW).}
\label{tab:Table4}
\includegraphics[width=0.95\textwidth]{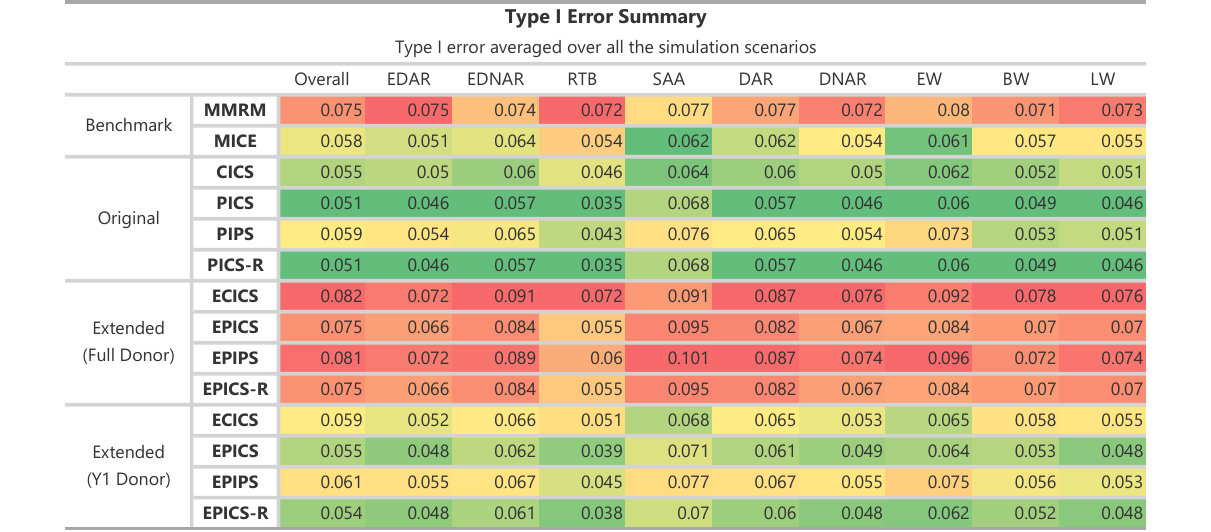}
\end{table}

\section{Discussion}\label{sec5}

This work was motivated by a structural gap in the sequential MI framework for treatment policy estimands \citep{ich_e9r1_2019,rubin_multiple_2004}: the framework assumes every subject contributes at least one post-baseline observation, an assumption violated by the baseline-only early dropout commonly observed in pSS trials. We proposed EPSMI, which combines donor-based reconstruction with an extended discontinuation-pattern indicator to allow all eight sequential MI models to be applied to the full randomized sample. Across the 24 scenarios examined, the Y1 Donor strategy proved broadly applicable and consistently strong-performing. More fundamentally, this contribution extends the applicability of an entire class of sequential MI models, removing a shared precondition that previously limited its use to trials without baseline-only dropout.

The performance gap between Full Donor and Y1 Donor has a clear explanation. Full Donor deterministically copies an entire trajectory from a single donor once prior to, rather than re-sampled within, each of the $m$ imputations, so donor-matching uncertainty is never propagated into the between-imputation variance component of Rubin's rules; under EDNAR, this compounds further as donor--recipient mismatch is carried forward across all three post-baseline visits. Y1 Donor avoids this by confining deterministic reconstruction to Y1---the most reliably matched visit---and leaving Y2 and Y3 to the pattern-extended sequential MI engine, which imputes them stochastically. We therefore recommend Y1 Donor combined with pattern-based extended models (EPICS, EPICS-R) as the primary analysis strategy; Full Donor may still be retained as a simple sensitivity analysis. Any application of EPSMI to real data should first confirm that the donor pool is sufficiently large and the caliper appropriately specified.

EPSMI can be positioned relative to two literatures. Relative to the sequential MI literature \citep{drury_estimation_2024}, EPSMI extends the applicability of an existing framework to a type of missingness it did not previously accommodate. Relative to the hot-deck and donor-based imputation literature \citep{fuller_hot_2005,andridge_review_2010}, its contribution lies in combining donor reconstruction with pattern-based sequential MI in a limited role: donor information fills only the first post-baseline visit, rather than substituting for formal statistical inference over the full trajectory.

This study has several limitations. First, donors were selected once rather than independently re-sampled within each imputation, a principal source of the variance underestimation observed for Full Donor. Second, donor matching relied on a simple Euclidean distance with a fixed caliper, without exploring more elaborate matching strategies. Third, the data-generating mechanism retained simplifying assumptions inherited from the underlying framework---shared on/off-treatment covariance, linear trajectories, and equal rates across arms---without a broader factorial sweep. Fourth, this study is simulation-based and has not been validated against real trial data.

Finally, the ICH E9 (R1) estimands framework emphasizes that the estimand should be defined independently of the estimation method \citep{ich_e9r1_2019}; a difficult missingness pattern should not implicitly narrow the population to which the estimand applies. As noted in Section 2.4, excluding baseline-only early dropouts biases the estimand toward a systematically selected subpopulation whenever early dropout is informative. EPSMI allows estimation to remain faithful to the treatment policy estimand as defined over the full randomized population, rather than requiring analysts to accept an implicitly narrowed estimand.

\section{Conclusion}\label{sec6}

This work identified and resolved a structural gap in the sequential MI framework for treatment policy estimands: its inapplicability, without ad hoc exclusion, to subjects who experience baseline-only early dropout and contribute no post-baseline outcome data. We proposed EPSMI, which combines covariate-matched donor reconstruction with an extension of the discontinuation-pattern indicator, allowing all eight existing sequential MI models to be applied to the full randomized sample rather than a reduced, non-early-dropout subpopulation. Across the factorial design of early-dropout, off-treatment, discontinuation, and withdrawal scenarios examined, Full Donor reconstruction is not recommended as a primary analysis strategy, as its deterministic, single-step reconstruction amplifies bias under informative early dropout and understates variance, leading to under-coverage and Type I error inflation in the most challenging scenarios; it may nonetheless be retained as a simple sensitivity analysis. Y1 Donor reconstruction, combined with pattern-based extended models such as EPICS and EPICS-R, instead proved consistently robust across nearly the full range of scenarios considered and is recommended as the primary analysis strategy when baseline-only early dropout is present. More broadly, EPSMI allows analysts to remain faithful to the treatment policy estimand as defined over the entire randomized population, without being forced, by the presence of baseline-only early dropout, to accept either an implicitly narrowed estimand or a loss of statistical efficiency---provided that the donor pool underlying the reconstruction is adequately validated for the trial at hand.

\section*{Author contributions}

All authors contributed to the conception and design of the study. Material preparation and data analysis were performed by Chen Zhang, Junyu Nie, and Kexuan Li. Chen Zhang wrote the first draft of the manuscript. All authors reviewed and critically revised the manuscript and approved the final version.

\section*{Acknowledgments}

The authors gratefully acknowledge Charlotte Baidoo for her sponsorship and support of this research. The authors also extend their appreciation to Ning Ding for her leadership and guidance, Junyu Nie for her mentorship and technical expertise, and Kexuan Li for his valuable discussions and insightful feedback. Their guidance, expertise, and support contributed significantly to the development and completion of this work.

\section*{Financial disclosure}

None reported.

\section*{Conflict of interest}

The authors declare no potential conflict of interest.

\bibliographystyle{plainnat}
\bibliography{references}

\end{document}